\documentclass[aps,prl,twocolumn,groupedaddress,amsmath,amssymb]{revtex4-1}
\usepackage{graphicx}  
\usepackage{dcolumn}   
\usepackage{bm}        
\usepackage{verbatim}   
\usepackage[latin1]{inputenc}
\usepackage{ragged2e}

\begin{document}

\title{The number radial coherent states for the generalized MICZ-Kepler problem}
\author{D. Ojeda-Guill\'en, M. Salazar-Ram\'irez}
\affiliation{Escuela Superior C\'omputo, Instituto Polit\'ecnico Nacional, Av. Juan de Dios B\'atiz esq. Av. Miguel Oth\'on de Mendiz\'abal, Col. Lindavista, Del. Gustavo A. Madero, C.P. 07738, M\'exico D. F., M\'exico.  }

\author{R.D. Mota}
\affiliation{Escuela Superior de Ingenier\'ia Mec\'anica y El\'ectrica, Unidad Culhuac\'an,
Instituto Polit\'ecnico Nacional,  Av. Santa Ana No. 1000, Col. San
Francisco Culhuac\'an, Del. Coyoac\'an, C.P. 04430,  M\'exico D. F., M\'exico.}

\date{\today}
\begin{abstract}
We study the radial part of the MICZ-Kepler problem in an algebraic way by using the $su(1,1)$ Lie algebra. We obtain the energy spectrum and the eigenfunctions of this problem from the $su(1,1)$ theory of unitary representations and the tilting transformation to the stationary Schr\"odinger equation. We construct the physical Perelomov number coherent states for this problem and compute some expectation values. Also, we obtain the time evolution of these coherent states.
 \end{abstract}

\maketitle

\section{1. Introduction}

The theory of coherent states arose in order to make comparisons between the classical and quantum theories. These states were introduced by Schr\"odinger \cite{Scrho} in 1926 as the most classical ones of the harmonic oscillator, i.e., those of minimal uncertainty which are not deformed when evolve in time along a classical trajectory. In 1963 Glauber \cite{Glau}, Klauder \cite{Klau1,Klau2} and Sudarshan \cite{Sudar} took up again the coherent states for application in quantum optics. Glauber defined the coherent states as the eigenstates of the annihilation operator of the harmonic oscillator and are related to those the Heisenberg-Weyl group.

The harmonic oscillator is not the only problem  for which we can construct coherent states. It has been shown that coherent states
can be built up for any problem from the dynamical properties of its associated group \cite{Perel2}. Some of these generalizations are reported in the work of Barut \cite{Barut} and Perelomov \cite{Perel}. In particular, the Barut and Girardello coherent states are a generalization of those introduced by Glauber, since were defined as the eigenstates of the annihilation operator $T_-$ of the $SU(1,1)$ group. Perelomov defined his coherent states
as the action of the displacement operator (constructed in terms of the algebra generators) on the ground state. The coherent states have been obtained successfully for many problems, reported in references \cite{Perel2,KlauLib, Gaz}.

Besides the harmonic oscillator, other problem for which it has been successfully constructed the coherent states is the hydrogen atom. In particular, Y. Gur and A. Mann \cite{Gur} obtained the radial Barut-Girardelllo coherent states related to the $su(1,1)$ Lie algebra for the harmonic oscillator and the hydrogen atom in arbitrary dimensions.

The $SU(1,1)$ Perelomov coherent states have been obtained in several works, among them is the work presented by Gerry and Kiefer \cite{Ger1}, who obtained the coherent states for the Coulomb problem. The wave packets they obtained evolve in a fictitious time variable proportional to the eccentric anomaly. These packets  consist of a superposition of the Coulomb Sturmian functions and do not disperse as they evolve although the package width changes periodically. Nieto \cite{Nieto} and his collaborators have obtained coherent states for several central potentials by using the method of minimum uncertainty, which  apparently are dispersed as they evolve in time

The $SU(1,1)$ Perelomov number coherent states were introduced by Gerry as the eigenfunctions of the degenerate parametric amplifier \cite{Ger2}. Moreover, it has been shown that the Perelomov number coherent states for the two-dimensional harmonic oscillator are useful in the study of the non-degenerate parametric amplifier \cite{Did1} and the problem of two coupled oscillators \cite{Did2}.

On the other hand, the MICZ-Kepler problem is the Kepler problem when the nucleus of this hypothetic hydrogen atom also carries a magnetic charge. This problem was independently discovered by McIntosh-Cisneros \cite{McI} and Zwanziger \cite{Zwan}. It has been shown that the MICZ-Kepler problem possesses a Runge-Lenz-type vector as constant of motion and the group $O(4)$ as symmetry group \cite{McI}. These facts reflect a great similarity with the Coulomb problem. Different generalizations for the MICZ-Kepler problem have been studied. For example, Meng \cite{Meng} has studied the MICZ-Kepler problem in all dimensions and Mardoyan \cite{Mard1,Mard2} has solved the Schr\"odinger equation for the generalized MICZ-Kepler Hamiltonian ($\hbar=m=c=1$)
\begin{equation}
H=\frac{1}{2}\left(-i\nabla-s\bold{A}\right)^2+\frac{s^2}{2r^2}-\frac{1}{r}+\frac{c_1}{r(r+z)}+\frac{c_2}{r(r-z)}.
\label{EQ1}
\end{equation}
In this Hamiltonian, $\bold A$ is the magnetic vector potential of a Dirac monopole,
\begin{equation}
{\bold A}=\frac{1}{r(r-z)}(y,-x,0),
\end{equation}
such that $\nabla\times {\bold A}=\frac{{\bold r}}{r}$, $r$ is the distance from the electron to the hydrogen nucleus, $s$ is the magnetic charge of the Dirac monopole which takes the values $s=0$, $\pm\frac{1}{2},\pm 1,\pm \frac{3}{2}$... and,  $c_1$  and $c_2$ are non-negative constants. Moreover, there are generalizations of the MICZ-Kepler systems on the three-dimensional sphere \cite{Grit} and on the hyperboloid \cite{Nerse}. The radial equation resulting from the Hamiltonian (\ref{EQ1}) has been solved
by supersymmetric quantum mechanics \cite{Giri}. Also, in Ref. \cite{INT} it has been constructed and $su(1,1)$ Lie algebra to solve the radial equation of (\ref{EQ1}) in an algebraic way.

The aim of this work is to study and construct the Perelomov radial number coherent states of the MICZ-Kepler problem by introducing an $su(1,1)$ Lie algebra and using the tilting transformation. In references \cite{Gur} and \cite{Ger1} we infer that the requirements to construct the coherent states are a set of radial operators independent of the energy, which close a Lie algebra, and the tilting transformation. At this point it is necessary to emphasize that although the $su(1,1)$ Lie algebra generators we employ in the present work are very similar to those we used in \cite{INT}, they are not the same. Those we used in our previous work
are reescaled with the energy, and therefore, are not useful to construct the coherent states.

This work  is organized as  follows. In Section $2$, we use some previous results to study the radial part of the MICZ-problem by using tree operators which close the $su(1,1)$ Lie algebra. We use the tilting transformation to obtain the energy spectrum and the radial functions. In Section $3$, we obtain the Perelomov number coherent states of this problem and calculate some expectation values using these states. In Section $4$ we obtain the most general form of the MICZ-Kepler problem coherent states, since our results include the time dependence. Finally, we give some concluding remarks.

\section{2. The Generalized MICZ-Kepler Problem in Spherical Basis and its radial $SU(1,1)$ dynamical group}
The stationary Schr\"odinger equation $H\Psi=E\Psi$ for the Hamiltonian (\ref{EQ1}) with $\Psi\equiv R(r)Z(\theta,\phi)$ in spherical coordinates $(r,\theta,\phi)$, can be reduced to the uncoupled differential equations \cite{Mard1,Mard1,Giri}
\begin{align}\nonumber
&\frac{1}{\sin\theta}\frac{\partial}{\partial \theta}\left(\sin\theta\frac{\partial Z}{\partial \theta}\right)+\frac{1}{4\cos^2\frac{\theta}{2}}\left(\frac{\partial^2}{\partial\phi^2}-4c_1\right)Z\\
&+\frac{1}{4\sin^2\frac{\theta}{2}}\left[\left(\frac{\partial}{\partial\phi}+2is\right)^2-4c_2\right]Z=-{\cal A}Z,\label{ANG}\\
&\frac{1}{r^2}\frac{d}{dr}\left(r^2\frac{d\overline{R}}{dr}\right)-\frac{{\cal A}}{r^2}\bar{R}+2\left(E+\frac{1}{r}\right)\overline{R}=0, \label{RAD}
\end{align}
where the quantized separation constant is given by
\begin{equation}
{\cal A}=\left(j+\frac{\delta_1+\delta_2}{2}\right)\left(j+\frac{\delta_1+\delta_2}{2}+1\right).
\label{QCON}
\end{equation}
The unnormalized square-integrable solutions for the equation (\ref{ANG}) are \cite{Mard1,Mard2}
\begin{align}\nonumber
Z^{(s)}_{jm}(\theta,\phi;\delta_1\delta_2)&=\left(\cos\frac{\theta}{2}\right)^{m_1} \left(\sin\frac{\theta}{2}\right)^{m_2}\times\\
&\times P^{(m_2,m_1)}_{j-m_+}(\cos\theta)e^{i(m-s)\phi},\label{ASOL}
\end{align}
where $m_1=|m-s|+\delta_1=\sqrt{(m-s)^2+4c_1}$,$m_2=|m+s|+\delta_2=\sqrt{(m+s)^2+4c_2}$, $m_+=(|m+s|+|m-s|)/2$ and $P^{(a,b)}_n$ are the Jacobi polynomials. The $z$-component of the total angular momentum $m$ and the total angular momentum $j$ take the quantized eigenvalues
\begin{align}
&m=-j,-j+1,...,j-1,j,\\
&j=\frac{|m+s|+|m-s|}{2}, \frac{|m+s|+|m-s|}{2}+1,....\label{j}
\end{align}
Notice that the Dirac quantization condition $s$ determines the values of $j$ and $m$. These two equations imply that $j$ and $m$ take integer or half-integer values depending on whether $s$ takes integer or half-integer values.

By substituting the quantized separation constant (\ref{QCON}) into equation (\ref{RAD}) we obtain
\begin{align}\nonumber
&\frac{1}{r^2}\frac{d}{dr}\left(r^2\frac{d\overline{R}}{dr}\right)-\frac{1}{r^2}\left(j+\frac{\delta_1+\delta_2}{2}\right)\left(j+\frac{\delta_1+\delta_2}{2}+1\right)\overline{R}\\
&+2\left(E+\frac{1}{r}\right)\overline{R}=0. \label{RAD2}
\end{align}
This equation can be written as
\begin{equation}
\left[-\frac{1}{2}\frac{d^2}{dr^2}-\frac{1}{r}\frac{d}{dr}+\frac{J(J+1)}{2r^2}-\frac{1}{r}-E\right]\overline{R}=0,
\end{equation}
where $\overline{R}$ is the physical wave function and $J=j+(\delta_1+\delta_2)/2$. We introduce the $su(1,1)$ Lie algebra generators for the radial part of the MICZ-Kepler problem as
\begin{align}\label{OPERTO}
T_0=&\frac{1}{2}\left[rp_r^2-\frac{J(J+1)}{2r}+r\right],\\\label{OPERT1}
T_1=&\frac{1}{2}\left[rp_r^2-\frac{J(J+1)}{2r}-r\right],\\\label{OPERT2}
T_2=&rp_r,
\end{align}
where
\begin{equation}
p_r=-\frac{1}{r}\left[\frac{\partial}{\partial r}r\right].
\end{equation}

By using equations (\ref{OPERTO}), (\ref{OPERT1}) and (\ref{OPERT2}) the eigenvalue problem satisfies
\begin{equation}\label{HAMa}
\overline{H}|\overline{R}\rangle=|\overline{R}\rangle,
\end{equation}
where
\begin{equation}
\overline{H}=\frac{1}{2}\left(T_0+T_1\right)-E\left(T_0-T_1\right),
\end{equation}
is a pseudo-Hamiltonian. We can remove the non-diagonal operator $T_1$ of equation (\ref{HAMa}) by applying the ``tilting'' transformation. Thus,
\begin{equation}
e^{-i\beta T_2}\overline{H}e^{i\beta T_2}e^{-iT_2}|\overline{R}\rangle=e^{-i\beta T_2}|\overline{R}\rangle,
\end{equation}
or
\begin{equation}\label{HA}
H|R\rangle=|R\rangle,
\end{equation}
with
\begin{equation}
|R\rangle=e^{-i\beta T_2}|\overline{R}\rangle,\hspace{3ex}\hbox{and}\hspace{3ex}H=e^{-i\beta T_2}\overline{H}e^{-i\beta T_2}\label{EIGEN}.
\end{equation}
 By using the Baker-Campbell-Hausdorff formula (\ref{Baker})
we obtain the following relationship
\begin{equation}
e^{-i\beta T_2}\left(T_0\pm T_1\right)e^{i\beta T_2}=e^{\pm\beta}\left(T_0\pm T_1\right).
\end{equation}
Thus, the tilted Hamiltonian $H$ can be expressed as
\begin{equation}\label{Ham1}
H=\frac{1}{2}e^{\beta}\left(T_0+T_1\right)-Ee^{-\beta}\left(T_0-T_1\right).
\end{equation}
In this equation $T_1$ vanishes if we choose the parameter $\beta$ as $\ln\left(-2E\right)^{1/2}$. Therefore, the tilted Hamiltonian becomes diagonal and can be written as
\begin{equation}
H=\left(-2E\right)^{1/2}T_0.\label{HAM2}
\end{equation}

On the other hand, the action of the Casimir operator (see equation (\ref{CAS}) Appendix ) on the radial function $|R\rangle$ is
\begin{equation}\label{CAS}
C|R\rangle=J(J+1)|R\rangle=k(k-1)|R\rangle.
\end{equation}
Thus, the relationship between the group number $k$ and the quantum number $J$ is $k=J+1$ or $k=-J$. Now, we just consider the positive solution since this leads to a unitary representation. From equations (\ref{HA}), (\ref{HAM2}) and (\ref{k0}) of Appendix, we obtain
\begin{equation}
\left(-2E\right)^{1/2}T_0|R\rangle=\left(-2E\right)^{1/2}(J+1+n)|R\rangle=|R\rangle.
\end{equation}
From this equality we obtain the energy spectrum of this problem
\begin{equation}
E=-\frac{1}{2(n+J+1)^2}.
 \end{equation}
By comparing this energy spectrum with the previously obtained in \cite{Mard1}, ($E=-1/(N+\delta_1+\delta_2)^2$) we obtain the relationship between the group number $n$ and the physical quantum numbers
\begin{equation}
n=N+\frac{\delta_1+\delta_2}{2}-(J+1).
\end{equation}

The physical states $|\overline{R}\rangle=|\overline{J,n}\rangle$ are obtained from the group states (Sturmian Basis)  $|J,n\rangle,\label{PState}$ as follows
\begin{equation}
|\overline{J,n}\rangle=C_{\scriptscriptstyle{n}}e^{i{\beta}T_2}|J,n\rangle,\label{PState}
\end{equation}
where $C_{\scriptscriptstyle {n}}$ is a normalization constant.
Notice that the physical states depend on energy, whereas the group states do not.
In order to find the normalization factor we require
\begin{align}
1=&\langle \overline{J,n}|r|\overline{J,n}\rangle\\
=&C_{\scriptscriptstyle{n}}^2e^{-\beta}\langle J,n|T_0-T_1|J,n\rangle,
\end{align}
from which we obtain $C_{\scriptscriptstyle {n}}=1/(J+n+1)$.

In Ref. \cite{Gur} and \cite{Ger1} it has been introduced the $su(1,1)$ radial Lie algebra and its Sturmian basis for its unitary irreducible representations to construct the coherent states for the Coulomb potential. Since the operators of equations (\ref{OPERTO})-(\ref{OPERT2}) of the $su(1,1)$ Lie algebra are a particular case of those of these works, we use that the group states in the configuration space are the radial Sturmian functions
\begin{equation}\label{CHI1}
R_{\scriptscriptstyle{nJ}}(r)=2\left[\frac{\Gamma(n+1)}{\Gamma(n+2J+2)}\right]e^{-r}(2r)^{J}L_{n}^{2J+1},
\end{equation}
where $L_{n}^{2J+1}$ are the associated Laguerre polynomials  \cite{Gur,Ger1}. These functions satisfy the orthonormality relation
\begin{equation}
\int R_{\scriptscriptstyle{nJ}}(r)\frac{1}{r}R_{\scriptscriptstyle{n'J'}}(r)r^2dr=\delta_{\scriptscriptstyle{n',n}}\delta_{\scriptscriptstyle{J,J'}}.
\end{equation}
The physical radial functions $\overline{R}_{\scriptscriptstyle{nJ}}(r)$ are obtained from equation (\ref{PState}) and the relationship
\begin{equation}
\exp\left[(\ln \lambda)r\partial r\right]f(r)=f\left(\lambda r\right).
\end{equation}
Thus, the physical radial functions of the MICZ-Kepler problem are
\begin{equation}\label{STUR1}
\overline{R}_{\scriptscriptstyle{nJ}}(r)=C_ne^{i\beta T_2}R_{\scriptscriptstyle{n,J}}(r)=\frac{1}{(n+J+1)^2}R_{\scriptscriptstyle{n,J}}\left[\frac{2r}{n}\right],
\end{equation}
and satisfy the orthonormality relation
\begin{equation}
\int \overline{R}_{\scriptscriptstyle{n,J}}(r)\overline{R}_{\scriptscriptstyle {n',J'}}(r)r^2dr=\delta_{\scriptscriptstyle {n,n'}}\delta_{\scriptscriptstyle {J,J'}}.
\end{equation}
All physical expectation values are calculated with the physical state $|\overline{J,n}\rangle$. Therefore it is important to remark that for some functions of $r$, $f(r)$, we have
\begin{equation}
\langle f(r)\rangle_{\scriptscriptstyle{nJ}}=\langle \overline{J,n}|rf(r)|\overline{J,n}\rangle
\end{equation}
with $r=T_0-T_1$.

\section{3. The SU(1,1) Perelomov Number Coherent State}

The $SU(1,1)$ Perelomov number coherent states are discussed in more detail in the Appendix. To obtain the number coherent states for the MICZ-Kepler problem we shall use the procedure described in the
reference \cite{Did1}. Thus, from the definition of the Perelomov number coherent states (see equation (\ref{PNCS})) and using the sums $(48.7.6)$ and $(48.7.8)$ of reference \cite{Hansen}, we obtain
\begin{align}\nonumber
\Psi_{J}\left(r,\zeta\right)\equiv &<r|\zeta,J,n\rangle=2(1-|\zeta|^2)^{J+1}e^{-r}e^{(2r\zeta)/(\zeta-1)}\\\nonumber
\times &(1-\zeta)^{-2J-2}\sqrt{\frac{\Gamma(n+1)}{\Gamma(2J+2+n)}}(2r)^J\\
\times & \left[\zeta^*(\sigma-1)\right]^n L_n^{2J+1}\left(\frac{2r\sigma}{(1-\zeta)(\sigma-1)}\right).
\end{align}
with $\sigma=\left(1-|\zeta|^2\right)/\zeta^*\left(1-\zeta\right)$. Furthermore, if we make $n=0$ in the previous result we obtain the standard $SU(1,1)$ Perelomov coherent states
\begin{align}\nonumber
\Psi_{\scriptscriptstyle{J}}(r,\zeta)
&=2\left(1-|\zeta|^2\right)^{J+1}\left[\Gamma\left(2J+2\right)\right]^{-1/2}\\
&\times e^{-r}\left(2r\right)^J\frac{e^{\frac{2r\zeta}{\zeta-1}}}{\left(1-\zeta\right)^{2J+2}}.\label{MICZCH}
\end{align}
From equation (\ref{HA}) the following relation is satisfied
\begin{equation}
\langle \zeta,J,n|H|\zeta,J,n\rangle=\sqrt{-2E}\langle \zeta,J,n|T_0|\zeta,J,n\rangle=1.
\end{equation}
Thus, using equation (\ref{cosh}) of Appendix we obtain that the energy depends on the group parameter $\theta$ as follows
\begin{equation}\label{Ener2}
E=-\frac{1}{2\left(J+n+1\right)^2\cosh^2\left(2|\xi|\right)}.
\end{equation}
Moreover, from equation (\ref{EIGEN}), the physical states of the MICZ-Kepler problem are given by
\begin{equation}
|\overline{\zeta,J,n}\rangle=Ce^{i\beta T_2}|\zeta,J,n\rangle,
\end{equation}
where $\beta=\ln(-2E)^{1/2}$ and $E$ is the continuous parameter of equation (\ref{Ener2}). Using equation (\ref{sinh}) and (\ref{cosh}) we obtain that the normalization requirement of the physical state leads to
\begin{align}\nonumber\label{CNOR1}
1&=\langle\overline{\zeta,J,n}|T_0-T_1|\overline{\zeta,J,n}\rangle\\
&=C^2e^{-\beta}\langle \zeta,J,n|T_0-T_1|\zeta,J,n\rangle\\\nonumber
&=C^2e^{-\beta}\left(J+n+1\right)\left[\cosh(2|\xi|)+\sinh(2|\xi|)\cos\varphi\right].
\end{align}
Thereby, the normalization constant is
\begin{equation}
C=\frac{\left(-2E\right)^{1/2}}{\left(J+n+1\right)^{1/2}\left[\cosh(2|\xi|)+\sinh(2|\xi|)\cos\varphi\right]^{1/2}}.
\end{equation}
Therefore, the physical wave function of the number coherent states for the MICZ-Kepler problem is
\begin{equation}
\bar{\Psi}_{\scriptscriptstyle{J,n}}\equiv\langle r|\overline{\xi,J,n}\rangle=C\left(-2E\right)^{1/2}\Psi_{\scriptscriptstyle{J,n}}\left(\sqrt{-2E}r,\xi\right).
\end{equation}

\section{4. TIME EVOLUTION OF THE $SU(1,1)$ NUMBER COHERENT STATE}
As it is well known from literature the time evolution operator for an arbitrary Hamiltonian is defined as $\mathcal{U}(t)=e^{-iHt/\hbar}$ \cite{Cohen}. Thus, if the Hamiltonian is proportional to $T_0$
\begin{equation}\label{UTEM}
\mathcal{U}(t)=e^{-i\gamma T_0t/\hbar}.
\end{equation}
In our particular case, for the tilted Hamiltonian $H$ and $\gamma=\left(-2E\right)^{1/2}$. The time evolution of the raising and lowering operators is obtained by applying a similarity transformation. Thus, from the BCH formula and the equations (\ref{UTEM}) and (\ref{opkmas}) we have
\begin{align}
T_+(t)=&\mathcal{U}^\dag(t)T_+\mathcal{U}(t)=T_+e^{i\gamma t/\hbar},\\
T_-(t)=&\mathcal{U}^\dag(t)T_-\mathcal{U}(t)=T_-e^{-i\gamma t/\hbar}
\end{align}
The same results are obtained by applying the Heisenberg equation. Thus, the time evolution of the Perelomov number coherent states is given by
\begin{equation}\label{PERET}
|\zeta(t), J,n\rangle =\mathcal{U}(t)|\zeta, J,n\rangle=\mathcal{U}(t)D(\xi)\mathcal{U}^\dag(t)\mathcal{U}(t)|J,n\rangle.
\end{equation}

From equation (\ref{k0}) of Appendix, the time evolution of the state $|J,n\rangle$ can be written as
\begin{equation}\label{evest1}
\mathcal{U}(t)|J,n\rangle=e^{-i\gamma(J+n)t/\hbar}|J,n\rangle.
 \end{equation}
The missing part $\mathcal{U}(t)D(\xi)\mathcal{U}^\dag(t)$ can be expressed as
\begin{align}\nonumber\label{opd1}
\mathcal{U}(t)D(\xi)\mathcal{U}^\dag(t)&=e^{\xi T_+(-t)-\xi^*T_-(-t)}\\
&=e^{\xi(-t)T_+ - \xi(-t)^*T_-},
\end{align}
hence, it follows that the time evolution of the displacement operator $D(\xi)$ is due to the time evolution of the complex  $\xi(t)=\xi e^{i\gamma t/\hbar}$ introduced in (\ref{opd1}). Moreover, the evolution of the displacement operator in its normal form is given by
\begin{equation}
D(\xi(t))=\mathcal{U}^\dag(t)D(\xi)\mathcal{U}(t)=\mathcal{U}^\dag(t)e^{\zeta T_+}e^{\eta T_0}e^{-\zeta^*T_-}\mathcal{U}(t),
\end{equation}
from which we can define the time dependent complex $\zeta(t)=\zeta e^{i\gamma t/\hbar}$ and obtain the time dependent normal form of the displacement operator $D(\xi)$ as
\begin{equation}\label{opedes}
D(\xi(t))=e^{\zeta(t) T_+}e^{\eta T_0}e^{-\zeta(t)^*T_-}.
\end{equation}

By means of equations (\ref{evest1}) and (\ref{opedes}), we obtain that the time dependent Perelomov number coherent state is
\begin{equation}
|\zeta(t),J,n\rangle=e^{-i\gamma(J+n)t/\hbar}e^{\zeta(-t)T_+}e^{\eta T_0}e^{-\zeta(-t)^*T_-}|J,n\rangle.
\end{equation}
With these results, we obtain that the time evolution of the number coherent state for the MICZ-Kepler problem in the configuration space is
\begin{align}\nonumber
\Psi_{J}\left(r,\zeta(t)\right)=&2(1-|\zeta|^2)^{J+1}e^{-i\gamma(J+n)t/\hbar}e^{-r}e^{(2r\zeta(-t))/(\zeta-1)}\\\nonumber
\times & (1-\zeta(-t))^{-2J-2}\sqrt{\frac{\Gamma(n+1)}{\Gamma(2J+2+n)}}(2r)^J\\
\times & \left[\zeta^*(-t)(\sigma-1)\right]^n L_n^{2J+1}\left(\frac{2r\sigma}{(1-\zeta(-t))(\sigma-1)}\right),
\end{align}
where $\gamma=\left(-2E\right)^{1/2}$, $n=N+\frac{\delta_1+\delta_2}{2}-(J+1)$ and $\zeta(-t)=\zeta e^{-i\gamma t/\hbar}$. This expression represents the most general radial number coherent states for the MICZ-Kepler problem, since it includes the time evolution. The particular case of equation (\ref{MICZCH}) is recovered by setting $t=0$ and $n=0$. Therefore, the time evolution of the physical number coherent state is given by
\begin{equation}
\bar{\Psi}_{\scriptscriptstyle{J,n}}\left(r,\zeta(t)\right)=\mathcal{N}\left(-2E\right)^{1/2}\Psi_{\scriptscriptstyle{J,n}}\left(\sqrt{-2E}r,\zeta(t)\right),
\end{equation}
where the normalization constant $\mathcal{N}$ is calculated from equation(\ref{CNOR1}) and is given by
\begin{equation}
\mathcal{N}=\frac{\left(-2E\right)^{1/2}}{\left(J+n+1\right)^{1/2}\left[\cosh(2|\xi|)+\sinh(2|\xi|)\cos(\varphi+\gamma t/\hbar)\right]^{1/2}}.
\end{equation}

\section{APPENDIX}

The generators $T_0$, $T_1$, and $T_2$  satisfy the  $su(1,1)$ Lie algebra,   which close the commutation relations
\begin{align}
\left[T_1,T_2\right]&=-iT_0,\\
\left[T_2,T_0\right]&=iT_1,\\
\left[T_0,T_1\right]&=iT_2,
\end{align}
or equivalently, in terms of the raising and lowering operators $T_{\pm}=T_1\pm iT_2$:
\begin{equation}\label{opkmas}
\left[T_0, T_\pm\right]=\pm T_\pm,\hspace{0.5cm}\left[T_-, T_+\right]=2T_0,
\end{equation}\label{relcom}
which act on the basis states  $|k,n\rangle$ (Sturmian bases) as
\begin{align}\label{k0}
T_0|k,n\rangle=&(n+k)|k,n\rangle,\\
T_+|k,n\rangle=&\left[(n+1)(n+2k)\right]^{1/2}|k,n+1\rangle,\\
T_-|k,n\rangle=&\left[n(n+2k-1)\right]^{1/2}|k,n-1\rangle.
\end{align}
The Casimir operator for the $su(1,1)$ lie algebra is given by
\begin{equation}\label{CAS}
C=T_0^2-T_1^2-T_2^2=T_0^2-\frac{1}{2}\left(T_+T_-+T_-T_+\right).
\end{equation}

The Perelomov number coherent states are constructed form action of the displacement operator $D(\xi)$
\begin{equation}
D(\xi)=\exp(\xi T_{+}-\xi^{*}T_{-}),
\end{equation}
on the basis states  $|k,n\rangle$. The properties
$T^{\dag}_{+}=T_{-}$ and $T^{\dag}_{-}=T_{+}$  allow to show that \cite{Perel2}
\begin{equation}
D^{\dag}(\xi)=\exp(\xi^{*} T_{-}-\xi T_{+})=D(-\xi).
\end{equation}
and the so called normal form of this operator
\begin{equation}
D(\xi)=\exp(\zeta T_{+})\exp(\eta T_{0})\exp(-\zeta^*T_{-})\label{normal}.
\end{equation}

Thus, by using the normal form of the displacement operator, the Perelomov number coherent states  are
\begin{align}\nonumber
|\zeta,k,n\rangle &=D(\xi)|k,n\rangle\\ &=\exp(\zeta T_{+})\exp(\eta
T_{3})\exp(-\zeta^* T_{-})|k,n\rangle,\label{defPCNS}
\end{align}
where $\xi=-\frac{1}{2}\tau e^{-i\varphi}$, $\zeta=-\tanh
(\frac{1}{2}\tau)e^{-i\varphi}$ and $\eta=-2\ln \cosh
|\xi|=\ln(1-|\zeta|^2)$ \cite{Perel2}. Thus, the Perelomov number coherent states in the Fock space are \cite{Did1}
\begin{align}\nonumber
|\zeta,k,n\rangle &=\sum_{s=0}^\infty\frac{\zeta^s}{s!}\sum_{j=0}^n\frac{(-\zeta^*)^j}{j!}e^{\eta(k+n-j)}\\
&\times\frac{\sqrt{\Gamma(2k+n)\Gamma(2k+n-j+s)}}{\Gamma(2k+n-j)}\nonumber\\
&\times\frac{\sqrt{\Gamma(n+1)\Gamma(n-j+s+1)}}{\Gamma(n-j+1)}|k,n-j+s\rangle.\label{PNCS}
\end{align}

Now, by using the Baker-Campbell-Hausdorff identity
\begin{align}\nonumber
e^{-A}Be^A=&B+\frac{1}{1!}[B,A]+\frac{1}{2!}[[B,A],A]+\\\label{Baker}
&\frac{1}{3!}[[[B,A],A],A]+...,
\end{align}
and equation (\ref{opkmas}), we can find the similarity transformations
$D^{\dag}(\xi)T_{+}D(\xi)$, $D^{\dag}(\xi)T_{-}D(\xi)$ and
$D^{\dag}(\xi)T_{0}D(\xi)$ of the $su(1,1)$ Lie algebra generators.
These results are given by
\begin{align}\label{simiK+}
D^{\dag}(\xi)T_{+}D(\xi) &=\frac{\xi^{*}}{|\xi|}\alpha T_{0}+\beta\left(T_{+}+\frac{\xi^{*}}{\xi}T_{-}\right)+T_{+},\\\label{simiK-}
D^{\dag}(\xi)T_{-}D(\xi) &=\frac{\xi}{|\xi|}\alpha T_{0}+\beta\left(T_{-}+\frac{\xi}{\xi^{*}}T_{+}\right)+T_{-},\\\label{simiK0}
D^{\dag}(\xi)T_{0}D(\xi)&=(2\beta+1) T_{0}+\frac{\alpha\xi}{2|\xi|}T_{+}+\frac{\alpha\xi^*}{2|\xi|}T_{-},
\end{align}
where $\alpha=\sinh(2|\xi|)$ and
$\beta=\frac{1}{2}\left[\cosh(2|\xi|)-1\right]$.

Moreover, the expectation values of the group generators $T_{\pm}, T_0$ in the Perelomov number coherent states
can be easily computed by using the similarity transformations, equations (\ref{simiK+})-(\ref{simiK0}).
Thus,
\begin{align}\label{sinh}
\langle \zeta,k,n|T_{\pm}|\zeta,k,n\rangle &=-e^{\pm{i\varphi}}\sinh(2|\xi|)(k+n),\\\label{cosh}
\langle \zeta,k,n|T_{0}|\zeta,k,n\rangle &=\cosh(2|\xi|)(k+n).
\end{align}

\section{5. CONCLUDING REMARKS}
We obtained the energy spectrum and the eigenfunctions for the radial part of the MICZ-Kepler problem in an algebraic way. In order to make the original Hamiltonian diagonal, we applied to it the tilting transformation.
By introducing the Sturmian functions (which is the basis for the unitary irreducible representations of the $su(1,1)$ Lie algebra) we computed the energy spectrum and the radial functions.
It is important to emphasize that the $su(1,1)$ Lie algebra generators that we employed in this work are not reescaled with the energy.

Also, by using the fact that the tilted Hamiltonian depends only on the third generator $T_0$, we were able to compute the time evolution of the physical radial number coherent state for this problem from the Sturmian basis.

\section{ACKNOWLEDGMENTS}
This work was in part supported by SNI-M\'exico, COFAA-IPN, EDI-IPN, EDD-IPN, SIP-IPN Project No. 20150935.


\begin{thebibliography}{99}
\bibitem{Scrho}     E. Schr\"odinger, Naturwiss. 14, 664 (1926).
\bibitem{Glau}      R.J. Glauber, Phys. Rev. 130, 2529 (1963).
\bibitem{Klau1}     J.R. Klauder, Ann. Phys. (N.Y.) 11, 123 (1960).
\bibitem{Klau2}     J.R. Klauder, J. Math. Phys. 4, 1055 (1963).
\bibitem{Sudar}     E.C.G. Sudarshan, Phys. Rev. Lett. 10, 227 (1963).
\bibitem{Perel2}    A.M. Perelomov, Generalized Coherent States and Their Applications, Springer, Berlin, 1986.
\bibitem{Barut}     A.O. Barut, L. Girardello, Commun. Math. Phys. 21, 41 (1971).
\bibitem{Perel}     A.M. Perelomov, Commun. Math. Phys. 26, 222 (1972).
\bibitem{KlauLib}   J.R. Klauder, B.S. Skagerstam, Coherent States-Applications in Physics and Mathematical Physics, World Scientific, Singapore, 1985.
\bibitem{Gaz}       J.P. Gazeau, Coherent States in Quantum Physics, Wiley-VCH, Berlin, 2009.
\bibitem{Gur}       Y. Gur, A. Mann, Phys. At. Nucl. 68, 1700 (2005).
\bibitem{Ger1}      C.C. Gerry, J. Kiefer, Phys. Rev.A 37, 665 (1988).
\bibitem{Nieto}     M.M. Nieto, Phys. Rev. D 22, 391 (1980).
\bibitem{Ger2}      C.C. Gerry, J. Kiefer, Phys. Rev.A 39, 3204 (1989).
\bibitem{Did1}      D. Ojeda-Guill\'en, R.D. Mota, V.D. Granados, J. Math. Phys. 55, 042109 (2014).
\bibitem{Did2}      D. Ojeda-Guill\'en, R.D. Mota, V.D. Granados, Commun. Theor. Phys. 64 (2015) 34.
\bibitem{McI}       H.V. McIntosh, A. Cisneros, J. Math. Phys. 11, 896 (1968).
\bibitem{Zwan}      D. Zwanziger, Phys. Rev. 176, 1480 (1968).
\bibitem{Meng}      G.W. Meng, J. Math. Phys. 48, 320 (2007).
\bibitem{Mard1}     L. Mardoyan, J. Math. Phys. 44, 4981 (2003).
\bibitem{Mard2}     L. Mardoyan, quant-ph/0310143v1 (2003).
\bibitem{Grit}      V.V. Gritsev, Y.A. Kurochkin, V.S. Otchik, J. Phys. A, Math. Gen. 33, 4903 (2000).
\bibitem{Nerse}     A. Nersessian, G. Pogosyan, Phys. Rev. A 63, 020103(R) (2001).
\bibitem{Giri}      P.R. Giri, Mod. Phys. Lett. A 23, 895 (2008).
\bibitem{INT}       M. Salazar-Ram{\'i}rez, D. Mart{\'i}nez, V.D. Granados, R.D. Mota, Int. J. theor. Phys. 49, 967(2010).
\bibitem{Hansen}    E.R. Hansen, A Table of Series and Products, Prentice-Hall, Englewood Cliffs, New Jersey, 1975.
\bibitem{Cohen}     C. Cohen-Tannoudji, B. Diu, F. Laloe, Quantum Mechanics, Wiley-VCH, Berlin, 1977.

\end{thebibliography}
\end{document}